\documentclass[twocolumn,prb,amsmath,amssymb,showpacs,superscriptaddress,floatfix]{revtex4}

\usepackage{graphicx}

 \DeclareMathOperator{\sgn}{\mathrm{sgn}}
 \DeclareMathOperator{\Real}{\mathrm{Re}}
 \DeclareMathOperator{\Imag}{\mathrm{Im}}

\begin{document}

\title{Nonmonotonic critical temperature in superconductor/ferromagnet bilayers}

 \author{Ya.~V.~Fominov}
 \email{fominov@landau.ac.ru}
 \affiliation{L.~D.~Landau Institute for Theoretical Physics RAS, 117940 Moscow, Russia}
 \affiliation{Department of Applied Physics, University of Twente, P.O.~Box~217, 7500 AE Enschede,
The~Netherlands}

 \author{N.~M.~Chtchelkatchev}
 \email{nms@landau.ac.ru}
 \affiliation{L.~D.~Landau Institute for Theoretical Physics RAS, 117940 Moscow, Russia}

 \author{A.~A.~Golubov}
 \email{a.golubov@tn.utwente.nl}
 \affiliation{Department of Applied Physics, University of Twente, P.O.~Box~217, 7500 AE Enschede,
The~Netherlands}

\date{17 February 2002}

\begin{abstract}
The critical temperature $T_c$ of a superconductor/ferromagnet (SF) bilayer can exhibit nonmonotonic
dependence on the thickness $d_f$ of the F layer. SF systems have been studied for a long time; according
to the experimental situation, the ``dirty'' limit is often considered which implies that the mean free
path in the layers is the second smallest spatial scale after the Fermi wavelength. However, all
calculations reported for the dirty limit were done with some additional assumptions, which can be
violated in actual experiments. Therefore, we develop a general method (to be exact, two independent
methods) for investigating $T_c$ as a function of the bilayer's parameters in the dirty case. Comparing
our theory with experiment, we obtain good agreement. In the general case, we observe three characteristic
types of $T_c(d_f)$ behavior: 1)~nonmonotonic decay of $T_c$ to a finite value exhibiting a minimum at
particular $d_f$, 2)~reentrant behavior, characterized by vanishing of $T_c$ in a certain interval of
$d_f$ and finite values otherwise, 3)~monotonic decay of $T_c$ and vanishing at finite $d_f$.
Qualitatively, the nonmonotonic behavior of $T_c(d_f)$ is explained by the interference of quasiparticles
in the F layer, which can be either constructive or destructive depending on the value of $d_f$.
\end{abstract}


\maketitle

\section{Introduction}

Superconductivity and ferromagnetism are two competing orders: while the former ``prefers'' an
antiparallel spin orientation of electrons in Cooper pairs, the latter forces the spins to align in
parallel. Therefore, their coexistence in one and the same material is possible only in a narrow interval
of parameters; hence the interplay between superconductivity and ferromagnetism is most conveniently
studied when the two interactions are spatially separated. In this case the coexistence of the two orders
is due to the proximity effect. Recently, much attention has been paid to properties of hybrid proximity
systems containing superconductors (S) and ferromagnets (F); new physical phenomena were observed and
predicted in these systems.\cite{Ryazanov,Kontos,Radovic,Tagirov_PRL,Buzdin_DOS,Nazarov_DOS} One of the
most striking effects in SF layered structures is highly nonmonotonic dependence of their critical
temperature $T_c$ on the thickness $d_f$ of the ferromagnetic layers. Experiments exploring this
nonmonotonic behavior were performed previously on SF multilayers such as Nb/Gd,\cite{Jiang}
Nb/Fe,\cite{Muhge} V/V-Fe,\cite{Aarts} and Pb/Fe,\cite{Lazar} but the results (and, in particular, the
comparison between the experiments and theories) were not conclusive.

To perform reliable experimental measurements of $T_c(d_f)$, it is essential to have $d_f$ large compared
to the interatomic distance; this situation can be achieved only in the limit of weak ferromagnets. Active
experimental investigations of SF bilayers and multilayers based on Cu-Ni dilute ferromagnetic alloys are
carried out by several groups.\cite{Ryazanov_new,Aarts_recent} In SF bilayers, they observed nonmonotonic
dependence $T_c(d_f)$. While the reason for this effect in multilayers can be the $0$--$\pi$
transition,\cite{Radovic} in a bilayer system with a single superconductor this mechanism is irrelevant,
and the cause of the effect is interference of quasiparticle, specific to SF structures.

In the present paper, motivated by the experiments of Refs.~\onlinecite{Ryazanov_new,Aarts_recent} we
theoretically study the critical temperature of SF bilayers. Previous theoretical investigations of $T_c$
in SF structures were concentrated on systems with thin or thick layers (compared to the corresponding
coherence lengths); with SF boundaries having very low or very high transparencies; the exchange energy
was often assumed to be much larger than the critical temperature; in addition, the methods for solving
the problem were usually
approximate.\cite{Radovic,Aarts,Lazar,Buzdin,Demler,Khusainov,Tagirov,Tagirov_PRL} The parameters of the
experiments of Refs.~\onlinecite{Ryazanov_new,Aarts_recent} do not correspond to any of the above limiting
cases. In the present paper we develop two approaches giving the opportunity to investigate not only the
limiting cases of parameters but also the intermediate region. Using our methods, we find different types
of nonmonotonic behavior of $T_c$ as a function of $d_f$, such as minimum of $T_c$ and even reentrant
superconductivity. Comparison of our theoretical predictions with the experimental data shows good
agreement.

A number of methods can be used for calculating $T_c$. When the critical temperature of the structure is
close to the critical temperature $T_{cs}$ of the superconductor without the ferromagnetic layer, the
Ginzburg--Landau (GL) theory applies. However, $T_c$ of SF bilayers may significantly deviate from
$T_{cs}$, therefore we choose a more general theory valid at arbitrary temperature --- the quasiclassical
approach.\cite{Usadel,LO,RS} Near $T_c$ the quasiclassical equations become linear. In the literature the
emerging problem is often treated with the help of the so-called ``single-mode''
approximation,\cite{Demler,Khusainov,Tagirov,Tagirov_PRL} which is argued to be qualitatively reasonable
in a wide region of parameters. However, this method is justified only in a specific region of parameters
which we find below. Moreover, below we show examples when this method fails even qualitatively. Thus
there is need for an exact solution of the linearized quasiclassical equations. The limiting case of
perfect boundaries and large exchange energy was treated by Radovi\'c \textit{et al.}\cite{Radovic}

Based on the progress achieved for calculation of $T_c$ in SN systems (where N denotes a nonmagnetic
normal material),\cite{Golubov} we develop a generalization of the single-mode approximation --- the
multi-mode method. Although this method seems to be exact, it is subtle to justify it rigorously.
Therefore we develop yet another approach (this time mathematically rigorous), which we call ``the method
of fundamental solution''. The models considered
previously\cite{Radovic,Aarts,Lazar,Buzdin,Demler,Khusainov,Tagirov,Tagirov_PRL} correspond to limiting
cases of our theory. A part of our results was briefly reported in Ref.~\onlinecite{JETPL}.

The paper is organized as follows. In Sec.~\ref{sec:model} we formulate the Usadel equations and the
corresponding boundary conditions. Section~\ref{sec:multi-mode} is devoted to the exact multi-mode method
for solving the general equations. An alternative exact method, the method of fundamental solution, is
presented in Sec.~\ref{sec:fund_sol}. In Sec.~\ref{sec:num_res} we describe results of our methods. In
Sec.~\ref{sec:discussion}, a qualitative explanation of our results is presented, applicability of the
results to multilayered structures is discussed, and the use of a complex diffusion constant is commented
upon. Conclusions are presented in Sec.~\ref{sec:conclusions}.
Appendixes~\ref{ap:sec:analytics},\ref{ap:sma} contain analytical results for limiting cases. Finally,
technical details of the calculations are given in Appendix~\ref{ap:sec:space dep expl}.

\section{Model}
\label{sec:model}

\begin{figure}
\includegraphics[width=40mm]{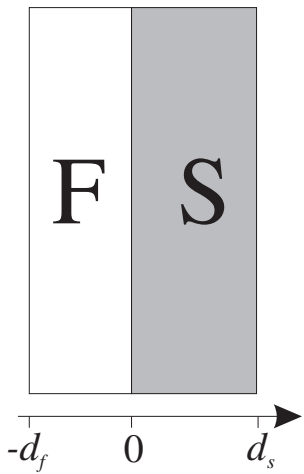}
\caption{\label{fig:fig1} SF bilayer. The F and S layers occupy the regions $-d_f<x<0$ and $0<x<d_s$,
respectively.}
\end{figure}
We assume that the dirty-limit conditions are fulfilled, and calculate the critical temperature of the
bilayer within the framework of the linearized Usadel equations for the S and F layers (the domain
$0<x<d_s$ is occupied by the S metal , $-d_f<x<0$ --- by the F metal, see Fig.~\ref{fig:fig1}). Near $T_c$
the normal Green function is $G=\sgn\omega_n$, and the Usadel equations for the anomalous function $F$
take the form
\begin{gather}
\label{U_1} \xi_s^2 \pi T_{cs} \frac{d^2F_s}{dx^2}- |\omega_n| F_s +\Delta=0,\quad 0<x<d_s;\\
\label{U_2} \xi_f^2 \pi T_{cs} \frac{d^2F_f}{dx^2} -(|\omega_n|+iE_\mathrm{ex} \sgn\omega_n) F_f=0, \\
-d_f<x<0; \nonumber\\
\label{U_3} \Delta\ln\frac{T_{cs}}{T} = \pi T \sum_{\omega_n} \left( \frac\Delta{|\omega_n|} -F_s \right)
\end{gather}
(the pairing potential $\Delta$ is nonzero only in the S part). Here $\xi_s=\sqrt{D_s/2\pi T_{cs}}$,
$\xi_f=\sqrt{D_f/2\pi T_{cs}}$ are the coherence lengths, while the diffusion constants can be expressed
via the Fermi velocity and the mean free path: $D=vl/3$; $\omega_n=\pi T(2n+1)$ with $n=0,\pm 1, \pm
2,\ldots$ are the Matsubara frequencies; $E_{\rm ex}$ is the exchange energy; and $T_{cs}$ is the critical
temperature of the S material. $F_{s(f)}$ denotes the function $F$ in the S(F) region. We use the system
of units in which Planck's and Boltzmann's constants equal unity, $\hbar=k_B=1$.

Equations (\ref{U_1})--(\ref{U_3}) must be supplemented with the boundary conditions at the outer surfaces
of the bilayer:
\begin{equation}
\frac{dF_s(d_s)}{dx}=\frac{dF_f(-d_f)}{dx}=0,
\end{equation}
as well as at the SF boundary:\cite{KL}
\begin{align}\label{bound_1}
& \xi_s\frac{dF_s(0)}{dx} =\gamma\xi_f \frac{dF_f(0)}{dx}, && \gamma =\frac{\rho_s\xi_s}{\rho_f\xi_f}, \\
& \xi_f\gamma_b \frac{dF_f(0)}{dx} = F_s(0)-F_f(0), && \gamma_b =\frac{R_b \mathcal{A}}{\rho_f\xi_f}.
\label{bound_2}
\end{align}
Here $\rho_s$, $\rho_f$ are the normal-state resistivities of the S and F metals, $R_b$ is the resistance
of the SF boundary, and ${\cal A}$ is its area. The Usadel equation in the F layer is readily solved:
\begin{gather}
F_f=C(\omega_n) \cosh\left( k_f [x+d_f]\right), \label{F_f} \\
k_f=\frac 1{\xi_f} \sqrt{\frac{|\omega_n|+iE_\mathrm{ex} \sgn\omega_n}{\pi T_{cs}}},\notag
\end{gather}
and the boundary condition at $x=0$ can be written in closed form with respect to $F_s$:
\begin{gather}
\xi_s \frac{dF_s(0)}{dx} =\frac\gamma{\gamma_b +B_f(\omega_n)} F_s(0), \label{bound_Phi} \\
B_f = \left[ k_f \xi_f \tanh (k_f d_f) \right]^{-1}.\notag
\end{gather}

This boundary condition is complex. In order to rewrite it in a real form, we do the usual trick and go
over to the functions
\begin{equation}
F^\pm =F(\omega_n)\pm F(-\omega_n).
\end{equation}
According to the Usadel equations (\ref{U_1})--(\ref{U_3}), there is the symmetry
$F(-\omega_n)=F^*(\omega_n)$ which implies that $F^+$ is real while $F^-$ is a purely imaginary function.

The symmetric properties of $F^+$ and $F^-$ with respect to $\omega_n$ are trivial, so we shall treat only
positive $\omega_n$. The self-consistency equation is expressed only via the symmetric function $F_s^+$:
\begin{equation} \label{self_cons}
\Delta\ln\frac{T_{cs}}T = \pi T \sum_{\omega_n>0} \left(\frac{2\Delta}{\omega_n}-F_s^+ \right),
\end{equation}
and the problem of determining $T_c$ can be formulated in a closed form with respect to $F_s^+$ as
follows. The Usadel equation for the antisymmetric function $F_s^-$ does not contain $\Delta$, hence it
can be solved analytically. After that we exclude $F_s^-$ from boundary condition (\ref{bound_Phi}) and
arrive at the effective boundary conditions for $F_s^+$:
\begin{equation} \label{bound_p}
\xi_s \frac{dF_s^+ (0)}{dx}= W(\omega_n) F_s^+ (0),\qquad\frac{dF_s^+ (d_s)}{dx}=0,
\end{equation}
where
\begin{gather}
W(\omega_n) =\gamma \frac{A_s (\gamma_b+\Real B_f)+ \gamma}{A_s |\gamma_b+B_f|^2 +\gamma
(\gamma_b+\Real B_f)}, \label{W_def} \\
A_s= k_s \xi_s \tanh (k_s d_s),\quad k_s=\frac 1{\xi_s} \sqrt{\frac{\omega_n}{\pi T_{cs}}} \notag.
\end{gather}
The self-consistency equation (\ref{self_cons}) and boundary conditions (\ref{bound_p})--(\ref{W_def}),
together with the Usadel equation for $F_s^+$:
\begin{equation}
\label{usadel_b} \xi_s^2 \pi T_{cs} \frac{d^2 F_s^+}{dx^2}- \omega_n F_s^+ +2\Delta=0
\end{equation}
will be used below for finding the critical temperature of the bilayer.

The problem can be solved analytically only in limiting cases (see Appendix~\ref{ap:sec:analytics}). In
the general case, one should use a numerical method, and below we propose two methods for solving the
problem exactly.

\section{Multi-mode method}
\label{sec:multi-mode}

\subsection{Starting point: the single-mode approximation and its applicability}
\label{sec:single-mode}

In the single-mode approximation (SMA) one seeks the solution of the problem
(\ref{self_cons})--(\ref{usadel_b}) in the form
\begin{gather}
F_s^+(x,\omega_n)=f(\omega_n)\cos\left(\Omega\frac{x-d_s}{\xi_s}\right), \label{Phi_sm} \\
\Delta(x)=\delta \cos\left(\Omega\frac{x-d_s}{\xi_s}\right). \label{Delta_sm}
\end{gather}
This anzatz automatically satisfies boundary condition (\ref{bound_p}) at $x=d_s$.

The Usadel equation (\ref{usadel_b}) yields
\begin{equation}
f(\omega_n)=\frac{2\delta}{\omega_n+ \Omega^2 \pi T_{cs}},
\end{equation}
then the self-consistency Eq. (\ref{self_cons}) takes the form ($\delta$ and $\Omega$ do not depend on
$\omega_n$)
\begin{equation} \label{Omega}
\ln\frac{T_{cs}}{T_c}=\psi\left(\frac 12+ \frac{\Omega^2}2 \frac{T_{cs}}{T_c}\right)- \psi\left(\frac
12\right),
\end{equation}
where $\psi$ is the digamma function.

Boundary condition (\ref{bound_p}) at $x=0$ yields
\begin{equation} \label{bound_s_mode}
\Omega\tan\left(\Omega \frac{d_s}{\xi_s} \right) = W(\omega_n).
\end{equation}
The critical temperature $T_c$ is determined by Eqs. (\ref{Omega}),(\ref{bound_s_mode}).

Although this method is popular, it is often used without pointing out the limits of its applicability. We
present the explicit formulation of the corresponding condition: the single-mode method is correct only if
the parameters are such that $W$ can be considered $\omega_n$-independent [because the left-hand side of
Eq. (\ref{bound_s_mode}) must be $\omega_n$-independent].\cite{Buzdin}

Appendix~\ref{ap:sma} demonstrates examples of the SMA validity and corresponding analytical results.

In one of experimentally relevant cases, $E_\mathrm{ex}/\pi T_{cs} > 1$, $d_f \sim \xi_f$, the SMA is
applicable if $\sqrt{E_\mathrm{ex}/\pi T_{cs}}\gg 1 /\gamma_b$ (see Appendix~\ref{ap:sma} for details).

\subsection{Inclusion of other modes}

The single-mode approximation implies that one takes the (only) real root $\Omega$ of Eq. (\ref{Omega}).
An exact (multi-mode) method for solving problem (\ref{self_cons})--(\ref{usadel_b}) is obtained if we
also take imaginary roots into account --- there is infinite number of these.\cite{Golubov}

Thus we seek the solution in the form
\begin{eqnarray}
F_s^+(x,\omega_n) &=& f_0(\omega_n)\cos\left(\Omega_0 \frac{x-d_s}{\xi_s} \right) \nonumber \\
&& + \sum_{m=1}^{\infty}f_m(\omega_n)\frac{\cosh\left(\Omega_m \frac{x-d_s}{\xi_s} \right)}
{\cosh\left(\Omega_m \frac{d_s}{\xi_s}\right)}, \label{Phi}
\end{eqnarray}
\begin{eqnarray}
\Delta(x) &=& \delta_0 \cos\left(\Omega_0\frac{x-d_s}{\xi_s}\right) \nonumber \\
&& + \sum_{m=1}^{\infty} \delta_m \frac{\cosh\left(\Omega_m\frac{x-d_s}{\xi_s}\right)}
{\cosh\left(\Omega_m\frac{d_s}{\xi_s}\right)}. \label{Delta}
\end{eqnarray}
(The normalizing denominators in the $\cosh$-terms have been introduced in order to increase accuracy of
numerical calculations.) This anzatz automatically satisfies boundary condition (\ref{bound_p}) at
$x=d_s$.

Substituting the anzatz [Eqs. (\ref{Phi})--(\ref{Delta})] into the Usadel equation (\ref{usadel_b}), we
obtain
\begin{align}
f_0(\omega_n) & =\frac{2\delta_0}{\omega_n+\Omega_0^2 \pi T_{cs}},\\
f_m(\omega_n) & =\frac{2\delta_m}{\omega_n-\Omega_m^2 \pi T_{cs}},\quad m=1,2,\ldots, \notag
\end{align}
then the parameters $\Omega$ are determined by the self-consistency equation (\ref{self_cons}) ($\delta$
and $\Omega$ do not depend on $\omega_n$):
\begin{align}
\ln\frac{T_{cs}}{T_c} & =\psi\left(\frac 12+ \frac{\Omega_0^2}2 \frac{T_{cs}}{T_c}\right)-
\psi\left(\frac 12\right), \label{Om} \\
\ln\frac{T_{cs}}{T_c} & =\psi\left(\frac 12- \frac{\Omega_m^2}2 \frac{T_{cs}}{T_c}\right)- \psi\left(\frac
12\right),\quad m=1,2,\ldots \notag
\end{align}
From Eqs. (\ref{Om}) and properties of the digamma function\cite{digamma} it follows that the parameters
$\Omega$ belong to the following intervals:
\begin{gather}
0< \Omega_0^2 < \frac 1{2\gamma_{_E}},\\
\frac{T_c}{T_{cs}} (2m-1) < \Omega_m^2 < \frac{T_c}{T_{cs}} (2m+1),\quad m=1,2,\ldots,\notag
\end{gather}
where $\gamma_{_E}\approx 1.78$ is Euler's constant.

Boundary condition (\ref{bound_p}) at $x=0$ yields the following equation for the amplitudes $\delta$:
\begin{eqnarray}
&& \delta_0 \frac{W(\omega_n) \cos\left(\Omega_0 d_s/\xi_s\right) -\Omega_0\sin\left( \Omega_0
d_s/\xi_s\right)}{\omega_n+\Omega_0^2 \pi T_{cs}} \notag \\
&& + \sum_{m=1}^\infty \delta_m \frac{W(\omega_n) + \Omega_m \tanh\left(\Omega_m
d_s/\xi_s\right)}{\omega_n-\Omega_m^2 \pi T_{cs}} =0. \label{bound_m_mode}
\end{eqnarray}
The critical temperature $T_c$ is determined by Eqs. (\ref{Om}) and the condition that Eq.
(\ref{bound_m_mode}) has a nontrivial ($\omega_n$-independent) solution with respect to $\delta$.

Numerically, we take a finite number of modes: $m=0,1,\ldots,M$. To take account of $\omega_n$-independence
of the solution, we write down Eq. (\ref{bound_m_mode}) at the Matsubara frequencies up to the $N$th
frequency: $n=0,1,\ldots,N$. Thus we arrive at the matrix equation $K_{nm}\delta_m=0$ with the following
matrix $\hat K$:
\begin{gather}
K_{n0} = \frac{W(\omega_n)\cos\left(\Omega_0 d_s/\xi_s \right) -\Omega_0\sin\left( \Omega_0
d_s/\xi_s\right)} {\omega_n/\pi T_{cs} +\Omega_0^2}, \notag \\
\label{K_def} K_{nm} = \frac{W(\omega_n) +\Omega_m \tanh\left(\Omega_m d_s/\xi_s\right)}{\omega_n/\pi
T_{cs} -\Omega_m^2}, \\
n=0,1,\ldots,N,\quad m=1,2,\ldots,M. \notag
\end{gather}
We take $M=N$, then the condition that Eq. (\ref{bound_m_mode}) has a nontrivial solution takes the form
\begin{equation} \label{K}
\det\hat K=0.
\end{equation}

Thus the critical temperature $T_c$ is determined as the largest solution of Eqs. (\ref{Om}),(\ref{K}).

\section{Method of fundamental solution}
\label{sec:fund_sol}

By definition, the fundamental solution $G(x,y;\omega_n)$ (which is also called the Green function) of
problem (\ref{bound_p})--(\ref{usadel_b}) satisfies the same equations, but with the delta-functional
``source'':\cite{Morse}
\begin{gather}
\xi_s^2 \pi T_{cs} \frac{d^2 G(x,y)}{dx^2}- \omega_n G(x,y) = - \delta(x-y),
\label{usadel_G} \\
\xi_s \frac{d G(0,y)}{dx}= W(\omega_n) G(0,y),\qquad\frac{d G(d_s,y)}{dx}=0.
\end{gather}
The fundamental solution can be expressed via solutions $v_1$, $v_2$ of Eq. (\ref{usadel_G}) without the
delta-function, satisfying the boundary conditions at $x=0$ and $x=d_s$, respectively:
\begin{eqnarray}
G(x,y;\omega_n) &=& \frac{k_s / \omega_n}{\sinh( k_s d_s) +(W/ k_s \xi_s) \cosh\left( k_s d_s\right)}
\notag \\
&& \times\left\{ \begin{aligned} v_1(x) v_2(y),\quad x\leqslant y\\ v_2(x) v_1(y),\quad y\leqslant x
\end{aligned} \right. , \label{G}
\end{eqnarray}
where
\begin{subequations}
\begin{align}
v_1(x) & =\cosh( k_s x) +(W/ k_s \xi_s) \sinh( k_s x),\\
v_2(x) & =\cosh\left( k_s [x-d_s] \right). \label{v_2}
\end{align}
\end{subequations}

Having found $G(x,y;\omega_n)$, we can write the solution of Eqs. (\ref{bound_p})--(\ref{usadel_b}) as
\begin{equation} \label{fsplus}
F_s^+ (x;\omega_n)=2\int_0^{d_s} G(x,y;\omega_n) \Delta(y) dy.
\end{equation}
Substituting this into the self-consistency equation (\ref{self_cons}), we obtain
\begin{multline}
\Delta(x) \ln\frac{T_{cs}}{T_c} \\
= 2\pi T_c \sum_{\omega_n>0} \left[
\frac{\Delta(x)}{\omega_n}-\int_0^{d_s} G(x,y;\omega_n) \Delta(y) dy \right]. \label{sc_Green}
\end{multline}
This equation can be expressed in an operator form: $\Delta\ln(T_{cs}/T_c)=\hat L\Delta$. Then the
condition that Eq. (\ref{sc_Green}) has a nontrivial solution with respect to $\Delta$ is expressed by the
equation
\begin{equation} \label{det}
\det\left(\hat L- \hat 1 \ln\frac{T_{cs}}{T_c} \right)=0.
\end{equation}
The critical temperature $T_c$ is determined as the largest solution of this equation.

Numerically, we put problem (\ref{sc_Green}),(\ref{det}) on a spatial grid, so that the linear operator
$\hat L$ becomes a finite matrix.

\section{Numerical results}
\label{sec:num_res}

In Secs. \ref{sec:multi-mode}, \ref{sec:fund_sol} we developed two methods for calculating the critical
temperature of a SF bilayer. Specifying parameters of the bilayer we can find the critical temperature
numerically. It can be checked that the multi-mode method and the method of fundamental solution yield
equivalent results. However, at small temperatures $T_c \ll T_{cs}$, the calculation time for the
multi-mode method increases. Indeed, the size of the matrix $\hat K$ [Eq. (\ref{K_def})] is determined by
the number $N$ of the maximum Matsubara frequency $\omega_N$, which must be much larger than the
characteristic energy $\pi T_{cs}$; hence $N\gg T_{cs}/T_c$. Therefore, at low temperatures we use the
method of fundamental solution.

\subsection{Comparison with experiment}

\begin{figure}
\includegraphics[width=85mm]{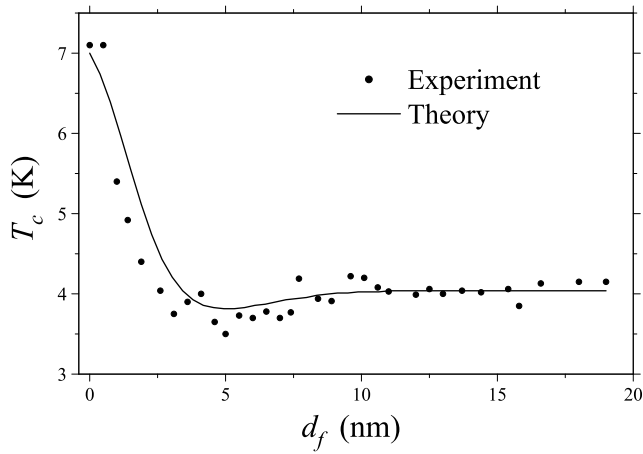}
\caption{\label{fig:ryazanov} Theoretical fit to the experimental data of Ref.~\onlinecite{Ryazanov_new}.
In the experiment, Nb was the superconductor (with $d_s=11$\,nm, $T_{cs}=7$\,K) and Cu$_{0.43}$Ni$_{0.57}$
was the weak ferromagnet. From our fit we estimate $E_\mathrm{ex}\approx 130$\,K and $\gamma_b\approx
0.3$.}
\end{figure}
Using our methods we fit the experimental data of Ref.~\onlinecite{Ryazanov_new}; the result is presented
in Fig.~\ref{fig:ryazanov}. Estimating the parameters $d_s=11$\,nm, $T_{cs}=7$\,K,
$\rho_s=7.5$\,$\mu\Omega$\,cm, $\xi_s=8.9$\,nm, $\rho_f=60$\,$\mu\Omega$\,cm, $\xi_f=7.6$\,nm,
$\gamma=0.15$ from the experiment,\cite{private_com} and fitting only $E_\mathrm{ex}$ and $\gamma_b$, we
find good agreement between our theoretical predictions and the experimental data.

The fitting procedure was the following: first, we determine $E_\mathrm{ex}\approx 130$\,K from the
position of the minimum of $T_c(d_f)$; second, we find $\gamma_b\approx 0.3$ from fitting the vertical
position of the curve.

The deviation of our curve from the experimental points is small; it is most pronounced in the region of
small $d_f$ corresponding to the initial decrease of $T_c$. This is not unexpected because, when $d_f$ is
of the order of a few nanometers, the thickness of the F film may vary significantly along the film (which
is not taken into account in our theory), and the thinnest films can even be formed by an array of islands
rather than by continuous material. At the same time, we emphasize that the minimum of $T_c$ takes place
at $d_f\approx 5$\,nm, when with good accuracy the F layer has uniform thickness.

\subsection{Various types of $T_c(d_f)$ behavior}

The experimental results discussed above represent only one possible type of $T_c(d_f)$ behavior. Now we
address the general case; we obtain different kinds of $T_c(d_f)$ curves depending on parameters of the
bilayer.

\begin{figure}
\includegraphics[width=85mm]{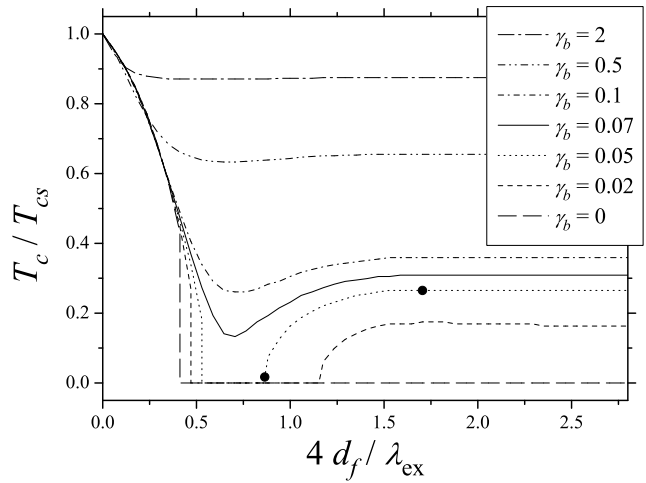}
\caption{\label{fig:tc_sf} Characteristic types of $T_c(d_f)$ behavior. The thickness of the F layer is
measured in units of the wavelength $\lambda_\mathrm{ex}$ defined in Eq. (\ref{lambda_ex}). The curves
correspond to different values of $\gamma_b$. The exchange energy is $E_\mathrm{ex}=150$\,K; the other
parameters are the same as in Fig.~\ref{fig:ryazanov}. One can distinguish three characteristic types of
$T_c(d_f)$ behavior: 1)~nonmonotonic decay to a finite $T_c$ with a minimum at particular $d_f$
($\gamma_{b}=2$; $0.5$; $0.1$; $0.07$), 2)~reentrant behavior ($\gamma_b=0.05$; $0.02$), 3)~monotonic
decay to $T_c=0$ at finite $d_f$ ($\gamma_b=0$). The bold points indicate the choice of parameter
corresponding to Fig.~\ref{fig:Delta}.}
\end{figure}

To illustrate, in Fig.~\ref{fig:tc_sf} we plot several curves for various values of $\gamma_b$ [we recall
that $\gamma_b\propto R_b$, where $R_b$ is the resistance of the SF interface in the normal state
--- see Eq. (\ref{bound_2})]. The exchange energy is $E_\mathrm{ex}=150$\,K; the other parameters are the
same as in Fig.~\ref{fig:ryazanov}.

We observe three characteristic types of $T_c(d_f)$ behavior: 1)~at large enough interface resistance,
$T_c$ decays nonmonotonically to a finite value exhibiting a minimum at a particular $d_f$, 2)~at moderate
interface resistance, $T_c$ demonstrates the reentrant behavior: it vanishes in a certain interval of
$d_f$, and is finite otherwise, 3)~at low enough interface resistance, $T_c$ decays monotonically
vanishing at finite $d_f$. A similar succession of $T_c(d_f)$ curves as in Fig.~\ref{fig:tc_sf} can be
obtained by tuning other parameters, e.g., the exchange energy $E_\mathrm{ex}$ or the normal resistances
of the layers (the parameter $\gamma$).

A common feature seen from Fig.~\ref{fig:tc_sf} is saturation of $T_c$ at large $d_f \gtrsim
\lambda_\mathrm{ex}$. This fact has a simple physical explanation: the suppression of superconductivity by
a dirty ferromagnet is only due to the effective F layer with thickness on the order of
$\lambda_\mathrm{ex}$, adjacent to the interface (this is the layer explored and ``felt'' by
quasiparticles entering from the S side due to the proximity effect).

\begin{figure}
\includegraphics[width=85mm]{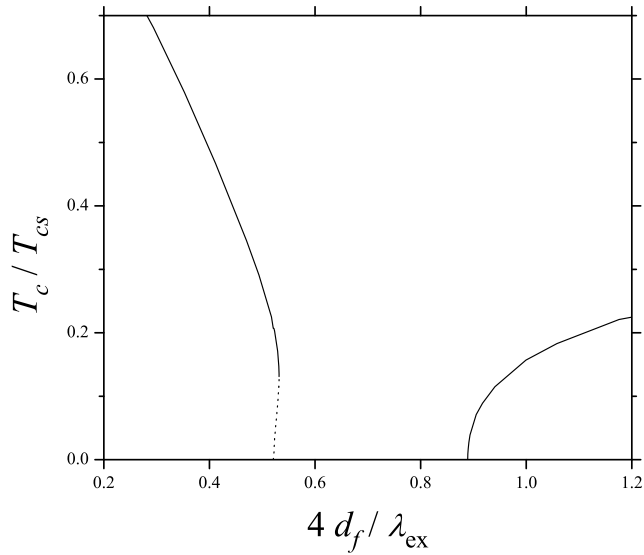}
\caption{\label{fig:zagib} Change of the phase transition's order. This phenomenon manifests itself as
discontinuity of $T_c(d_f)$: the critical temperature jumps to zero abruptly without taking intermediate
values. Formally, $T_c$ becomes a double-valued function, but the smaller solution is physically unstable
(dotted curve). For illustration we have chosen the curve from Fig.~\ref{fig:tc_sf} corresponding to
$\gamma_b=0.05$.}
\end{figure}

It was shown by Radovi\'{c} \textit{et al.}\cite{Radovic} that the order of the phase transition may
change in short-periodic SF superlattices, becoming the first order. We also observe this feature in the
curves of types 2) and 3) mentioned above. This phenomenon manifests itself as discontinuity of
$T_c(d_f)$: the critical temperature jumps to zero abruptly without taking intermediate values (see
Figs.~\ref{fig:tc_sf},\ref{fig:zagib}). Formally, $T_c$ becomes a double-valued function, but the smaller
solution is physically unstable (dotted curve in Fig.~\ref{fig:zagib}).

An interesting problem is determination of the tricritical point where the order of the phase transition
changes. The corresponding result for homogeneous bulk superconductors with internal exchange field was
obtained a long time ago in the framework of the Ginzburg--Landau theory.\cite{Sarma} However, the
generalization to the case when the GL theory is not valid is a subtle problem which has not yet been
solved. We note that the equations used in Refs.~\onlinecite{Radovic,Khusainov} were applied beyond their
applicability range because they are GL results valid only when $T_c$ is close to $T_{cs}$.

\subsection{Comparison between single- and multi-mode methods}

\begin{figure}
\includegraphics[width=85mm]{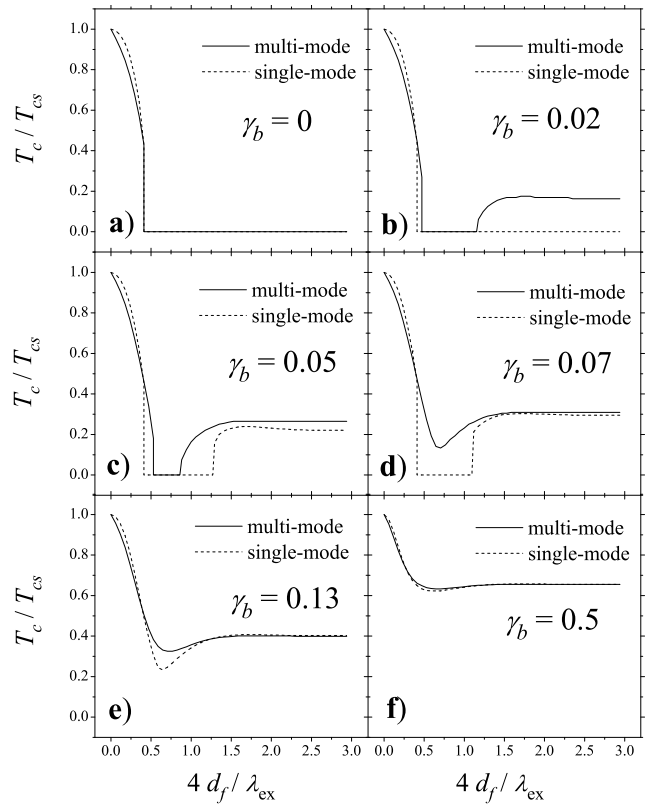}
\caption{\label{fig:compare} Comparison between single- and multi-mode methods. The parameters are the
same as in Fig.~\ref{fig:tc_sf}. Generally speaking, the results of the single-mode and multi-mode (exact)
methods are quantitatively and even qualitatively different: b), c), d), and e). However, sometimes the
results are close: a) and f). Thus the single-mode approximation can be used for quick estimates, but
reliable results should be obtained by one of the exact (multi-mode or fundamental-solution) techniques.}
\end{figure}

A popular method widely used in the literature for calculating the critical temperature of SF bi- and
multi-layers is the single-mode approximation. The condition of its validity was formulated in
Sec.~\ref{sec:single-mode}. However, this approximation is often used for arbitrary system's parameters.
Using the methods developed in Secs.~\ref{sec:multi-mode},\ref{sec:fund_sol}, we can check the actual
accuracy of the single-mode approximation. The results are presented in Fig.~\ref{fig:compare}.

We conclude that although at some parameters the results of the single-mode and multi-mode (exact) methods
are close (Figs.~\ref{fig:compare} a,f), in the general case they are quantitatively and even
qualitatively different [Figs.~\ref{fig:compare} b,c,d,e --- these cases correspond to the most nontrivial
$T_c(d_f)$ behavior]. Thus to obtain reliable results one should use one of the exact (multi-mode or
fundamental-solution) techniques.

\subsection{Spatial dependence of the order parameter}

The proximity effect in the SF bilayer is characterized by the spatial behavior or the order parameter,
which can be chosen as
\begin{equation} \label{order_parameter}
F(x,\tau=0) = T \sum_{\omega_n} F(x,\omega_n),
\end{equation}
where $\tau$ denotes the imaginary time [in the S metal $F(x,\tau=0)\propto \Delta(x)$]. This function is
real due to the symmetry relation $F(-\omega_n)=F^*(\omega_n)$.

\begin{figure}
 \includegraphics[width=85mm]{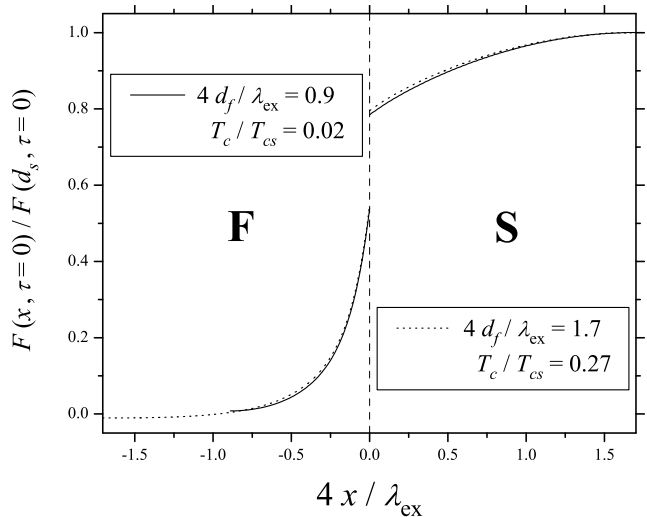}
 \caption{\label{fig:Delta} Spatial dependence of the order
parameter normalized by its value at the outer surface of the S layer. Two cases are shown differing by
the thickness of the F layer $d_f$ (and by the corresponding $T_c$) at $\gamma_b=0.05$. The other
parameters are the same as in Fig.~\ref{fig:tc_sf}, where the chosen cases are indicated by the bold
points. Although the critical temperatures differ by more than the order of magnitude, the normalized
order parameters are very close to each other, which means that the value of $T_c$ has almost no effect on
the shape of $F(x,\tau=0)$. The jump at the SF interface is due to its finite resistance. With an increase
of $d_f$ the order parameter starts to oscillate, changing its sign (this can be seen for the dotted
curve, although negative values of the order parameter have very small amplitudes).}
\end{figure}
We illustrate this dependence in Fig.~\ref{fig:Delta}, which shows two cases differing by the thickness of
the F layer $d_f$ (and by the corresponding $T_c$). Although the critical temperatures differ by more than
the order of magnitude, the normalized order parameters are very close to each other, which means that the
value of $T_c$ has almost no effect on the shape of $F(x,\tau=0)$. Details of the calculation are presented
in Appendix~\ref{ap:sec:space dep expl}.

Another feature seen from Fig.~\ref{fig:Delta} is that the order parameter in the F layer changes its sign
when the thickness of the F layer increases (this feature can be seen for the dotted curve, although
negative values of the order parameter have very small amplitudes). We discuss this oscillating behavior
in the next section.

\section{Discussion}
\label{sec:discussion}

\subsection{Qualitative explanation of the nonmonotonic $T_c(d_f)$ behavior}

The thickness of the F layer at which the minimum of $T_c(d_f)$ occurs, can be estimated from qualitative
arguments based on the interference of quasiparticles in the ferromagnet.

\begin{figure}
\includegraphics[width=85mm]{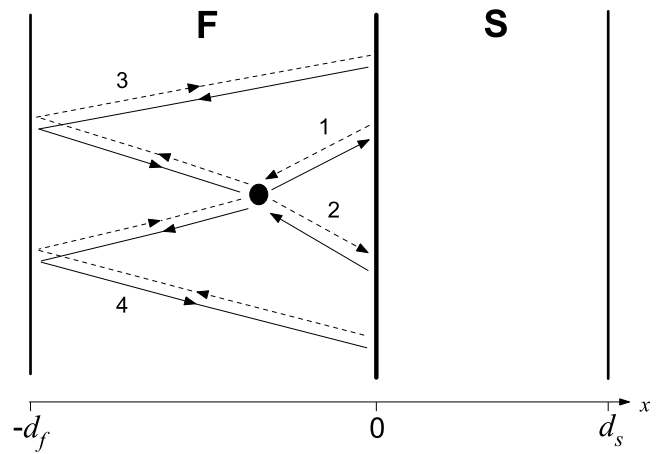}
\caption{\label{fig:intrfrnc} Four types of trajectories contributing (in the sense of Feynman's path
integral) to the anomalous wave function of correlated quasiparticles in the ferromagnetic region. The
solid lines correspond to electrons, the dashed lines --- to holes; the arrows indicate the direction of
the velocity.}
\end{figure}

Let us consider a point $x$ inside the F layer. According to Feynman's interpretation of quantum
mechanics,\cite{Feynman} the quasiparticle wave function may be represented as a sum of wave amplitudes
over all classical trajectories; the wave amplitude for a given trajectory is equal to $\exp(iS)$, where
$S$ is the classical action along this trajectory. We are interested in an \textit{anomalous} wave
function of correlated quasiparticles, which characterizes superconductivity; this function is equivalent
to the anomalous Green function $F(x)$. To obtain this wave function we must sum over trajectories that
(i) start and end at the point $x$, (ii) change the type of the quasiparticle (i.e., convert an electron
into a hole, or vice versa). There are four kinds of trajectories that should be taken into account (see
Fig.~\ref{fig:intrfrnc}).
Two of them (denoted 1 and 2) start in the direction toward the SF interface (as an electron and as a
hole), experience the Andreev reflection, and return to the point $x$. The other two trajectories (denoted
3 and 4) start in the direction away from the interface, experience normal reflection at the outer surface
of the F layer, move toward the SF interface, experience the Andreev reflection there, and finally return
to the point $x$. The main contribution is given by the trajectories normal to the interface. The
corresponding actions are
\begin{gather}
S_1=-Qx-\alpha,\\
S_2=Qx-\alpha, \\
S_3=-Q(2d_f+x)-\alpha,\\
S_4=Q(2d_f+x)-\alpha
\end{gather}
(note that $x<0$), where $Q$ is the difference between the wave numbers of the electron and the hole, and
$\alpha=\arccos(E/\Delta)$ is the phase of the Andreev reflection. To make our arguments more clear, we
assume that the ferromagnet is strong, the SF interface is ideal, and consider the clean limit first: then
$Q=k_e-k_h=\sqrt{2m(E+E_\mathrm{ex}+\mu)}- \sqrt{2m(-E-E_\mathrm{ex}+\mu)}\approx 2E_\mathrm{ex}/v$, where
$E$ is the quasiparticle energy, $\mu$ is the Fermi energy, and $v$ is the Fermi velocity. Thus the
anomalous wave function of the quasiparticles is
\begin{equation}
F(x)\propto \sum_{n=1}^4 \exp(iS_n)\propto \cos(Qd_f) \cos\left(Q[d_f+x] \right).
\end{equation}

The suppression of $T_c$ by the ferromagnet is determined by the value of the wave function at the SF
interface: $F(0)\propto \cos^2 (Qd_f)$. The minimum of $T_c$ corresponds to the minimum value of $F(0)$
which is achieved at $d_f=\pi/2Q$. In the dirty limit the above expression for $Q$ is replaced by
\begin{equation} \label{lambda_ex}
Q = \sqrt{\frac{E_\mathrm{ex}}{D_f}}\equiv \frac{2\pi}{\lambda_\mathrm{ex}}
\end{equation}
(here we have defined the wavelength of the oscillations $\lambda_\mathrm{ex}$); hence the minimum of
$T_c(d_f)$ takes place at
\begin{equation}
d_f^\mathrm{(min)} = \frac\pi 2 \sqrt{\frac{D_f}{E_\mathrm{ex}}} =\frac{\lambda_\mathrm{ex}}4.
\end{equation}
For the bilayer of Ref.~\onlinecite{Ryazanov_new} we obtain $d_f^\mathrm{(min)}\approx 7$\,nm, whereas the
experimental value is $5$\,nm (Fig.~\ref{fig:ryazanov}); thus our qualitative estimate is reasonable.

The arguments given above seem to yield not only the minimum but rather a succession of minima and maxima.
However, numerically we obtain either a single minimum or a minimum followed by a weak maximum
(Fig.~\ref{fig:tc_sf}). The reason for this is that actually the anomalous wave function not only
oscillates in the ferromagnetic layer but also decays exponentially, which makes the amplitude of the
subsequent oscillations almost invisible.

Finally, we note that our arguments concerning oscillations of $F(x)$ also apply to a half-infinite
ferromagnet, where we should take into account only the trajectories 1 and 2 (see
Fig.~\ref{fig:intrfrnc}). This yields $F(x)\propto\cos(Qx)$ (another qualitative explanation of this
result can be found, for example, in Ref.~\onlinecite{Demler}).

\subsection{Multilayered structures}

The methods developed and the results obtained in this paper apply directly to more complicated symmetric
multilayered structures in the $0$-state such as SFS and FSF trilayers, SFIFS and FSISF systems (I denotes
an arbitrary potential barrier), and SF superlattices. In such systems an SF bilayer can be considered as
a unit cell, and joining together the solutions of the Usadel equations in each bilayer we obtain the
solution for the whole system (for more details see Sec.~VIII of Ref.~\onlinecite{FF}).

Our methods can be generalized to take account of possible superconductive and/or magnetic $\pi$-states
(when $\Delta$ and/or $E_\mathrm{ex}$ may change their signs from layer to layer). In this case the system
cannot be equivalently separated into a set of bilayers. Mathematically, this means that the solutions of
the Usadel equations lose their purely cosine form [see Eqs. (\ref{F_f}), (\ref{Phi_sm}), (\ref{Delta_sm}),
(\ref{Phi}), (\ref{Delta}), (\ref{v_2})] acquiring a sine part as well.

\subsection{Complex diffusion constant?}

Finally, we comment on Refs.~\onlinecite{Tagirov,Tagirov_PRL,Khusainov,Buzdin_new}, where the authors
considered (in the vicinity of $T_c$) diffusion equations with a \textit{complex} diffusion constant $D_f$
for the F part of the structure. This implies small complex corrections to $D_f$ over $E_\mathrm{ex} \tau
\ll 1$ in the Usadel equations ($\tau$ is the time of the mean free path). However, we disagree with this
method for the following reason: although the complex $D_f$ can indeed be formally obtained in the course
of the standard derivation of the Usadel equations\cite{Usadel} from the Eilenberger
ones\cite{Eilenberger} by expanding over the spherical harmonics, one can check that the higher harmonics
neglected in the derivation have the same order of magnitude as the retained complex correction to $D_f$.
Hence the complexity of $D_f$ in the context of the Usadel equations is the excess of accuracy. Below we
present our arguments.

We give a brief derivation of the Usadel equations showing how the complex diffusion constant can be
obtained and why this result cannot be trusted. In the ``quasi-one-dimensional'' geometry (which means
that the parameters vary only as a function of $x$) the linearized Eilenberger equation in the presence of
disorder and the exchange field has the form
\begin{equation} \label{eilenberger}
\frac{v\cos\theta}2 \frac d{dx} F + \left( \omega_n+\frac 1{2\tau} +iE_\mathrm{ex} \right)F=\Delta
+\frac{\left< F\right>}{2\tau},
\end{equation}
where, for simplicity, we assume a positive Matsubara frequency $\omega_n>0$, and $\theta$ is the angle
between the $x$ axis and the direction of the Fermi velocity $\mathbf v$, while $\left<\dots\right>$
denotes angular averaging over the spherical angles. The disorder is characterized by the time of the mean
free path $\tau$ and the mean free path $l$ (to be used below). In the dirty limit the anomalous Green
function $F$ is nearly isotropic. However, to obtain the Usadel equation for the isotropic part of $F$, we
must also take into account the next term from the full Legendre polynomial expansion:
\begin{align}
F(x,\omega_n,\theta) &= \sum_{k=0}^\infty F_k(x,\omega_n) P_k(\cos\theta) \notag \\
& \approx F_0(x,\omega_n) +F_1(x,\omega_n)\cos\theta. \label{legendre}
\end{align}
Here we have neglected the harmonics with $k\geqslant 2$ assuming them small; we shall check this
assumption later.

Averaging Eq. (\ref{eilenberger}) over the spherical angles first directly and second --- after being
multiplied by $\cos\theta$, we arrive at
\begin{gather}
\frac v6 \frac d{dx} F_1 + \left(\omega_n+iE_\mathrm{ex} \right)F_0 =\Delta, \label{aux0}\\
\frac v2 \frac d{dx} F_0 + \left(\omega_n+\frac 1{2\tau} +iE_\mathrm{ex} \right)F_1=0. \label{aux}
\end{gather}
Equation (\ref{aux}) yields
\begin{equation} \label{F_1}
F_1 = -\left( \frac l{1+2\omega_n\tau+2iE_\mathrm{ex}\tau} \right) \frac d{dx} F_0,
\end{equation}
then Eq. (\ref{aux0}) leads to
\begin{gather} \label{Usadel}
\frac D2  \frac{d^2}{dx^2} F_0 - \left(\omega_n+ i E_\mathrm{ex} \right)F_0 + \Delta =0,\\
D = \frac{vl/3}{1+2\omega_n\tau+2iE_\mathrm{ex}\tau}. \notag
\end{gather}

Now we must check that the assumption $|F_1/F_0| \ll 1$, $|F_2/F_1| \ll 1$, etc. that we used is indeed
satisfied. From Eq. (\ref{F_1}) we obtain
\begin{equation}
\left| \frac{F_1}{F_0} \right| \sim \frac{l/L}{\max(1,2\omega_n\tau, 2 E_\mathrm{ex}\tau)},
\end{equation}
where $L$ is the characteristic space scale on which $F_0$ varies. According to the Usadel equation
(\ref{Usadel}), it is given by
\begin{equation}
L \sim \frac l{\sqrt{\max(1,2\omega_n\tau, 2 E_\mathrm{ex}\tau) \max( 2\omega_n\tau, 2
E_\mathrm{ex}\tau)}},
\end{equation}
and the condition of the Usadel equation's validity is written as
\begin{equation} \label{condition}
\left| \frac{F_1}{F_0} \right| \sim \sqrt{\frac{\max(2\omega_n\tau,2E_\mathrm{ex}\tau)} {\max(1,
2\omega_n\tau, 2E_\mathrm{ex}\tau)}} \ll 1
\end{equation}
[similarly, we can also keep the term with $k=2$ in series (\ref{legendre}), which yields $|F_2/F_1| \sim
|F_1/F_0|$, etc.].

Finally, condition (\ref{condition}) takes the form
\begin{equation} \label{condition_final}
2\pi T_{cs} \tau \ll 1,\qquad 2E_\mathrm{ex}\tau \ll 1
\end{equation}
(we have taken into account that the characteristic energy is $\omega_n\sim \pi T_{cs}$).

Now we can analyze our results. If condition (\ref{condition_final}) is satisfied and the Usadel equation
is valid, the neglected angular harmonics have the relative order of magnitude $|F_2/F_0|\sim \max(2\pi
T_{cs} \tau,2E_\mathrm{ex}\tau)$; hence we cannot retain the terms of the same order in the diffusion
constant [see Eq. (\ref{Usadel})], and we should use the standard expression $D=vl/3$.

\section{Conclusions}
\label{sec:conclusions}

In the present paper we have developed two methods for calculating the critical temperature of a SF
bilayer as a function of its parameters (the thicknesses and material parameters of the layers, the
quality of the interface). The multi-mode method is a generalization of the corresponding approach
developed in Ref.~\onlinecite{Golubov} for SN systems. However, the rigorous justification of this method
is not clear. Therefore, we propose yet another approach --- the method of fundamental solution, which is
mathematically rigorous. The results demonstrate that the two methods are equivalent; however, at low
temperatures (compared to $T_{cs}$) the accuracy requirements are stricter for the multi-mode method, and
the method of fundamental solution is preferable. Comparing our method with experiment we obtain good
agreement.

In the general case, we observe three characteristic types of $T_c(d_f)$ behavior: 1)~nonmonotonic decay
of $T_c$ to a finite value exhibiting a minimum at particular $d_f$, 2)~reentrant behavior, characterized
by vanishing of $T_c$ in a certain interval of $d_f$ and finite values otherwise, 3)~monotonic decay of
$T_c$ and vanishing at finite $d_f$. Qualitatively, the nonmonotonic behavior of $T_c(d_f)$ is explained
by interference of quasiparticles in the F layer, which can be either constructive or destructive
depending on the value of $d_f$.

Using the developed methods we have checked the accuracy of the widely used single-mode approximation. We
conclude that although at some parameters the results of the single-mode and exact methods are close, in
the general case they are quantitatively and even qualitatively different. Thus, to obtain reliable
results one should use one of the exact (multi-mode or fundamental-solution) techniques.

The spatial dependence of the order parameter (at the transition point) is shown to be almost insensitive
to the value of $T_c$.

The methods developed and the results obtained in this paper apply directly to more complicated symmetric
multilayered structures in the $0$-state such as SFS and FSF trilayers, SFIFS and FSISF systems, and SF
superlattices. Our methods can be generalized to take account of possible superconductive and/or magnetic
$\pi$-states (when $\Delta$ and/or $E_\mathrm{ex}$ may change their signs from layer to layer).

We argue that the use of the complex diffusion constant in the Usadel equation is the excess of accuracy.

In several limiting cases, $T_c$ is considered analytically.

\begin{acknowledgments}
We thank V.~V.~Ryazanov and M.~V.~Feigel'man for stimulating discussions. We are especially indebted to
V.~V.~Ryazanov for communicating the experimental result of his group to us prior to the detailed
publication. We are also grateful to J.~Aarts, A.~I.~Buzdin, M.~Yu.~Kupriyanov, Yu.~Oreg, and
L.~R.~Tagirov for useful comments. Ya.V.F. acknowledges financial support from the Russian Foundation for
Basic Research (RFBR) under project No. 01-02-17759, and from Forschungszentrum J\"ulich (Landau
Scholarship). The research of N.M.C. was supported by the RFBR (projects Nos. 01-02-06230 and
00-02-16617), by Forschungszentrum J\"ulich (Landau Scholarship), by the Netherlands Organization for
Scientific Research (NWO), by the Einstein Center, and by the Swiss National Foundation.
\end{acknowledgments}

\appendix

\section{Analytical results for a thin S layer}
\label{ap:sec:analytics}

(i) When $d_s \ll \xi_s$ and $E_\mathrm{ex} \gg \pi T_{cs}$, problem (\ref{self_cons})--(\ref{usadel_b})
can be solved analytically. The first of the above conditions implies that $\Delta$ can be considered
constant, and $F^+$ weakly depends on the spatial coordinate; so $F^+(x,\omega_n)=2\Delta/\omega_n +
A(\omega_n)\cosh(k_s[x-d_s])$. The boundary conditions determine the coefficient $A$; as a result
\begin{equation} \label{thin_S_F^+}
F^+(\omega_n)\equiv F^+(x=0,\omega_n)=\frac{2\Delta}{\omega_n} \left[
\frac{A_s(\omega_n)}{A_s(\omega_n)+W(\omega_n)}\right],
\end{equation}
where $k_s$, $A_s$, and $W$ are defined in Eq. (\ref{W_def}). Finally, the self-consistency equation for
$T_c$ takes the form
\begin{equation} \label{thin_S}
\ln\frac{T_{cs}}{T_c} = \Real \psi\left( \frac 12 +\frac\gamma 2 \frac{\xi_s}{d_s} \frac 1{\gamma_b+B_f}
\frac{T_{cs}}{T_c} \right) -\psi\left( \frac 12 \right),
\end{equation}
where $B_f$ does not depend on $\omega_n$ due to the condition $E_\mathrm{ex} \gg \pi T_{cs}$:
\begin{equation}
B_f=\left[ k_f \xi_f \tanh (k_f d_f) \right]^{-1},\qquad k_f=\frac 1 {\xi_f}\sqrt{\frac{iE_\mathrm{ex}}{\pi
T_{cs}}}.
\end{equation}

(ii) If the F layer is also thin, $d_f\ll \sqrt{D_f/2 E_\mathrm{ex}}$, Eq. (\ref{thin_S}) is further
simplified:
\begin{equation}
\ln\frac{T_{cs}}{T_c} = \Real \psi\left( \frac 12 + \frac{\tau_f}{\tau_s} \left[ \frac 1{-i+\tau_f
E_\mathrm{ex}} \right] \frac{E_\mathrm{ex}}{2\pi T_c} \right) -\psi\left( \frac 12 \right),
\end{equation}
where $\tau_s$, $\tau_f$ are defined similarly to Ref.~\onlinecite{FF}:
\begin{equation} \label{tau}
\tau_s = \frac{2 d_s R_b \mathcal{A}}{\rho_s D_s},\qquad \tau_f = \frac{2 d_f R_b \mathcal{A}}{\rho_f D_f},
\end{equation}
and have the physical meaning of the escape time from the corresponding layer. They are related to the
quantities $\gamma$, $\gamma_b$ used in the body of the paper as
\begin{equation} \label{tau_gamma}
\tau_s =\frac{\gamma_b}\gamma \frac 1{\pi T_{cs}} \frac{d_s}{\xi_s},\qquad \tau_f =\gamma_b \frac 1{\pi
T_{cs}} \frac{d_f}{\xi_f}.
\end{equation}

(iii) If the S layer is thin, $d_s\ll \xi_s$, and the SF interface is opaque, $\gamma_b\to\infty$, the
critical temperature of the bilayer only slightly deviates from $T_{cs}$. In this limit Eq.
(\ref{thin_S_F^+}) applies with $W=\gamma/\gamma_b \ll 1$, and we finally obtain:
\begin{equation} \label{Tc_opaque}
T_c = T_{cs}-\frac\pi{4\tau_s}.
\end{equation}
Interestingly, characteristics of the F layer ($d_f$, $E_\mathrm{ex}$, etc.) do not enter the formula. In
particular, this formula is valid for an SN bilayer\cite{McMillan,T_c_McMillan} (where N is a nonmagnetic
normal material, $E_\mathrm{ex}=0$) because Eq. (\ref{Tc_opaque}) was obtained without any assumptions
about the value of the exchange energy.

\subsection{Transparent interface}

When both layers are very thin [$d_s\ll \sqrt{D_s / 2 \omega_{_D}}$, $d_f \ll \min(\sqrt{D_f /
2\omega_{_D}}, \sqrt{D_f/2 E_\mathrm{ex}})$, with $\omega_{_D}$ the Debye energy of the S metal] and the
interface is transparent, the bilayer is equivalent to a homogeneous superconducting layer with internal
exchange field. This layer is described by effective parameters: the pairing potential
$\Delta^{(\mathrm{eff})}$, the exchange field $E_\mathrm{ex}^{(\mathrm{eff})}$, and the pairing constant
$\lambda^{(\mathrm{eff})}$. In this subsection we develop the ideas of Ref.~\onlinecite{Bergeret-Efetov},
demonstrate a simple derivation of this description, and find the limits of its applicability.

The Usadel equations (\ref{U_1}),(\ref{U_2}) for the two layers can be written as a single equation:
\begin{multline}
\label{U_thin}
\frac{D_f\theta(-x)+D_s\theta(x)}{2}\frac{d^2 F}{dx^2}\\
-|\omega_n|F-iE_\mathrm{ex}\sgn(\omega_n) \theta(-x)F+\Delta\theta(x)=0,
\end{multline}
where $\theta$ is the Heaviside function [$\theta(x>0)=1$, $\theta(x<0)=0$]. The self-consistency equation
(\ref{U_3}) can be rewritten as
\begin{equation}
\Delta(x)=\lambda \theta(x) \pi T\sum_{\omega_n} F(x,\omega_n),
\end{equation}
where $\lambda$ is the pairing constant.

First, we consider the ideal SF interface: $\gamma_b=0$ [see Eq. (\ref{bound_2})], then $F(x)$ is
continuous at the interface and nearly constant across the \textit{whole} bilayer, i.e., $F_s(x)\approx
F_f(x)=F$. Applying the integral operator to Eq. (\ref{U_thin}):
\begin{equation}
\frac{\nu_f}{\nu_sd_s+\nu_f d_f}\int_{-d_f}^0 dx + \frac{\nu_s}{\nu_sd_s+\nu_f d_f}\int_0^{d_s}dx
\end{equation}
(here $\nu$ is the normal-metal density of states), and cancelling gradient terms due to the boundary
condition (\ref{bound_1}), we obtain the equations describing a homogeneous layer:
\begin{gather}
-|\omega_n| F(\omega_n)-i E_\mathrm{ex}^{(\mathrm{eff})} \sgn(\omega_n) F(\omega_n)+
\Delta^{(\mathrm{eff})}=0, \label{homo} \\
\Delta^{(\mathrm{eff})}=\lambda^{(\mathrm{eff})} \pi T\sum_{\omega_n} F(\omega_n),
\end{gather}
with the effective parameters (see also Ref.~\onlinecite{Bergeret-Efetov}):
\begin{gather}
\label{U_eff} E_\mathrm{ex}^{(\mathrm{eff})} = \frac{\tau_f}{\tau_s+\tau_f} E_\mathrm{ex},\qquad
\Delta^{(\mathrm{eff})} = \frac{\tau_s}{\tau_s+\tau_f} \Delta,\\
\lambda^{(\mathrm{eff})} = \frac{\tau_s}{\tau_s+\tau_f} \lambda,\qquad T_{cs}^{(\mathrm{eff})} =
\frac{\gamma_{_E}}\pi 2\omega_{_D} \exp\left( -\frac 1{\lambda^{(\mathrm{eff})}} \right),\notag
\end{gather}
where $\gamma_{_E}$ is Euler's constant and $T_{cs}^{(\mathrm{eff})}$ is the critical temperature of the
layer in the absence of ferromagnetism (i.e., at $E_\mathrm{ex}^{(\mathrm{eff})} = 0$). The critical
temperature is determined by the equation
\begin{equation}
\ln\frac{T_{cs}^{(\mathrm{eff})}}{T_c} = \Real \psi\left( \frac 12 +
i\frac{E_\mathrm{ex}^{(\mathrm{eff})}}{2\pi T_c}  \right) -\psi\left( \frac 12 \right).
\end{equation}

Actually, the description in terms of effective parameters (\ref{U_eff}) is applicable at an arbitrary
temperature (i.e., when the Usadel equations are nonlinear) and has a clear physical interpretation: the
superconducting ($\Delta$, $\lambda$) and ferromagnetic ($E_\mathrm{ex}$) parameters are renormalized
according to the part of time spent by quasiparticles in the corresponding layer. This physical picture is
based on interpretation of $\tau$ as escape times, which we present in the next subsection.

Now we discuss the applicability of the above description for a nonideal interface ($\gamma_b\neq 0$). In
this case $F$ is nearly constant in each layer, but these constants are different: $F_s(x)\approx F_s +
C_s (x-d_s)^2$, $F_f(x)\approx F_f+C_f(x+d_f)^2$, where $|F_s|\gg |C_s| d_s^2$ and $|F_f|\gg |C_f| d_f^2$.
Using the Usadel equation (\ref{U_thin}) and the boundary conditions (\ref{bound_1}),(\ref{bound_2}), we
find the difference $\delta F\equiv F_s-F_f$:
\begin{equation}
\delta F =\frac\Delta{\displaystyle \frac 1{\tau_s} +|\omega_n| \left[ 1+\frac 1{\tau_f \left( \left|
\omega_n \right|+i E_\mathrm{ex} \sgn\omega_n \right) } \right] }.
\end{equation}
Finally, the homogeneous description is valid when $|\delta F/F|\ll 1$ [with $F$ determined by Eq.
(\ref{homo})], which yields:
\begin{equation}
\max(E_\mathrm{ex}, \omega_{_D}) \max(\tau_s,\tau_f)\ll 1
\end{equation}
(here $\omega_n \sim \omega_{_D}$ has been taken as the largest characteristic energy scale in the
quasi-homogeneous bilayer).

\subsection{Interpretation of $\tau$ as escape times}

The quantities $\tau_s$, $\tau_f$ introduced in Eq. (\ref{tau}) may be interpreted as escape times from the
corresponding layers. The arguments go as follows.

If the layers are thin, then the diffusion inside the layers is ``fast'' and the escape time from a layer
is determined by the interface resistance. The time of penetration through a layer or the interface is
determined by the corresponding resistance: $R_{s(f)}$ or $R_b$, hence the diffusion is ``fast'' if
$R_{s(f)} \ll R_b$.

Let us use the detailed balance approach, and consider an interval of energy $dE$. In the S layer, the
charge in this interval is $Q_s=e \nu_s dE \mathcal{A} d_s$. Let us define the escape time from the S
layer $t_s$, so that the current from S to F is equal to $Q_s/t_s$. On the other hand, this current can be
written as $dE/eR_b$, hence
\begin{equation}
\frac{Q_s}{t_s}=\frac{dE}{eR_b},
\end{equation}
and we immediately obtain
\begin{equation}
t_s = \frac{d_s R_b \mathcal{A}}{\rho_s D_s}.
\end{equation}
Similarly, we obtain the expression for the escape time from the F layer $t_f$. As a result, the relations
between the quantities $\tau$ defined in Eq. (\ref{tau}) and the escape times $t$ are simply
\begin{equation}
\tau_s=2t_s,\qquad \tau_f=2t_f.
\end{equation}

Microscopic expressions for the escape times may be obtained using the Sharvin formula for the interface
resistance. Assuming, for definiteness, that the Fermi velocity is smaller in the S metal, $v_s<v_f$, we
obtain
\begin{equation} \label{R_int}
R_b = \frac{\pi r_b}{e^2 \nu_s v_s \mathcal{A}},
\end{equation}
and consequently
\begin{equation} \label{t_micro}
t_s=\pi \frac{d_s}{v_s} r_b,\qquad t_f=\pi \frac{v_f d_f}{v_s^2} r_b,
\end{equation}
where $r_b$ is the inverse transparency of one channel. The asymmetry in these expressions stems from our
assumption $v_s<v_f$. In the opposite case the indices $s$ and $f$ in Eqs. (\ref{R_int}),(\ref{t_micro})
should be interchanged.

\section{Applicability of the single-mode approximation}
\label{ap:sma}

As pointed out in Sec.~\ref{sec:single-mode}, the single-mode approximation (SMA) is applicable only if
the parameters are such that $W$ [see Eq. (\ref{W_def})] can be considered $\omega_n$-independent. An
example is the case when $\gamma_b \gg |B_f|$, hence $W= \gamma/ \gamma_b$.

The condition $\gamma_b \gg |B_f|$ can be written in a simpler form; to this end we should estimate
$|B_f|$. We introduce the real and imaginary parts of $k_f$: $k_f= k_f' +ik_f''$, and note that
$k_f'>k_f''$. Then using the properties of the trigonometric functions and the estimate $\tanh x \sim
\min(1,x)$ we obtain
\begin{equation} \label{est1}
|B_f| \sim \left[ k_f' \xi_f \tanh (k_f' d_f) \right]^{-1},
\end{equation}
and finally cast the condition $\gamma_b \gg |B_f|$ into the form
\begin{equation} \label{condition_SMA}
\frac 1{\gamma_b} \ll \min\left\{ \sqrt{\max\left( \frac{T_c}{T_{cs}}, \frac{E_\mathrm{ex}}{\pi T_{cs}}
\right)}; \frac{d_f}{\xi_f} \max\left( \frac{T_c}{T_{cs}}, \frac{E_\mathrm{ex}}{\pi T_{cs}} \right)
\right\},
\end{equation}
where the ratio $T_c /T_{cs}$ originates from $\omega_n / \pi T_{cs}$ with $\omega_n \sim \pi T_c$ as the
characteristic energy scale in the bilayer.

If condition (\ref{condition_SMA}) is satisfied, then the SMA is valid and $T_c$ is determined by the
equations
\begin{gather}
\ln\frac{T_{cs}}{T_c}=\psi\left(\frac 12+ \frac{\Omega^2}2 \frac{T_{cs}}{T_c}\right)- \psi\left(\frac
12\right),  \label{sc}\\
\Omega \tan\left(\Omega \frac{d_s}{\xi_s} \right) = \frac\gamma{\gamma_b}. \label{bc}
\end{gather}

These equations can be further simplified in two limiting cases which we consider below.

(1) $\displaystyle \frac\gamma{\gamma_b} \frac{d_s}{\xi_s} \ll 1$:

in this case Eq. (\ref{bc}) yields $\Omega^2 =\frac\gamma{\gamma_b} \frac{\xi_s}{d_s}$, and Eq. (\ref{sc})
takes the form
\begin{equation} \label{sma_res1}
\ln\frac{T_{cs}}{T_c}=\psi\left(\frac 12+ \frac 12 \frac\gamma{\gamma_b} \frac{\xi_s}{d_s}
\frac{T_{cs}}{T_c} \right)- \psi\left(\frac 12\right),
\end{equation}
which reproduces the $\gamma_b \gg |B_f|$ limit of Eq. (\ref{thin_S}).

(2) $\displaystyle \frac\gamma{\gamma_b} \frac{d_s}{\xi_s} \gg 1$:

in this case Eq. (\ref{bc}) yields $\Omega \frac{d_s}{\xi_s} =\frac\pi 2$, and Eq. (\ref{sc}) takes the
form
\begin{equation} \label{sma_res2}
\ln\frac{T_{cs}}{T_c}=\psi\left(\frac 12+ \frac{\pi^2}8 \left[ \frac{\xi_s}{d_s} \right]^2
\frac{T_{cs}}{T_c} \right)- \psi\left(\frac 12\right).
\end{equation}

Equations (\ref{sc})--(\ref{sma_res2}) can be used for calculating the critical temperature $T_c$ and the
critical thickness of the S layer $d_s^{(\mathrm{cr})}$ below which the superconductivity in the SF bilayer
vanishes (i.e., $T_c=0$).

\subsection{Results for the critical temperature}

In the limit when $T_c$ is close to $T_{cs}$, Eqs. (\ref{sma_res1}),(\ref{sma_res2}) yield
\begin{equation} \label{T_c_weak_dev}
T_c =T_{cs} \left( 1-\frac{\pi^2}4 \frac\gamma{\gamma_b} \frac{\xi_s}{d_s}\right) \qquad\text{if }
\frac\gamma{\gamma_b} \ll \min\left( \frac{d_s}{\xi_s},\frac{\xi_s}{d_s} \right),
\end{equation}
and
\begin{equation}
T_c=T_{cs} \left[1 - \left( \frac{\pi^2}4 \frac{\xi_s}{d_s} \right)^2 \right]\qquad\text{if }
\frac{d_s}{\xi_s} \gg \max\left( 1, \frac{\gamma_b}\gamma \right).
\end{equation}
Using relations (\ref{tau_gamma}) one can check that result (\ref{T_c_weak_dev}) is equivalent to Eq.
(\ref{Tc_opaque}).

\subsection{Results for the critical thickness}

The critical thickness of the S layer $d_s^{(\mathrm{cr})}$ is defined as the thickness below which there
is no superconductivity in the SF bilayer: $T_c ( d_s^{(\mathrm{cr})} )=0$. When $T_c \to 0$, Eq.
(\ref{sc}) yields $\Omega =1/\sqrt{2\gamma_{_E}}$ (where $\gamma_{_E} \approx 1.78$ is Euler's constant),
and Eq. (\ref{bc}) takes the form
\begin{equation}
\frac 1{\sqrt{2\gamma_{_E}}} \tan\left(\frac 1{\sqrt{2\gamma_{_E}}} \frac{d_s^{(\mathrm{cr})}}{\xi_s}
\right) = \frac\gamma{\gamma_b}.
\end{equation}
Explicit results for $d_s^{(\mathrm{cr})}$ can be obtained in limiting cases:
\begin{equation}
\frac{d_s^{(\mathrm{cr})}}{\xi_s} =2\gamma_{_E} \frac\gamma{\gamma_b} \qquad\text{if }
\frac\gamma{\gamma_b} \frac{d_s}{\xi_s} \ll 1,
\end{equation}
and
\begin{equation}
\frac{d_s^{(\mathrm{cr})}}{\xi_s} =\pi \sqrt{\frac{\gamma_{_E}}2} \qquad\text{if } \frac\gamma{\gamma_b}
\frac{d_s}{\xi_s} \gg 1.
\end{equation}

\section{Spatial dependence of the order parameter}
\label{ap:sec:space dep expl}

According to the self-consistency equation, in the S layer the order parameter $F(x,\tau=0)$ is
proportional to $\Delta(x)$:
\begin{equation}
F_s(x,\tau=0) = \frac{\Delta(x)}{\pi\lambda},
\end{equation}
where $\lambda$ is the pairing constant which can be expressed via the Debye energy:
\begin{equation}
\lambda^{-1} =\ln\left( \frac{2\gamma_{_E} \omega_{_D}}{\pi T_{cs}} \right).
\end{equation}
The pairing potential $\Delta(x)$ can be found as the eigenvector of the matrix $\hat L- \hat 1
\ln(T_{cs}/T_c)$ [see Eq. (\ref{det})], corresponding to the zero eigenvalue.

After that we can express $F(x,\tau=0)$ in the F layer via $\Delta(x)$ in the superconductor. The Green
function $F_f(x,\omega_n)$ in the F layer is given by Eq. (\ref{F_f}), with $C(\omega_n)$ found from the
boundary conditions:
\begin{equation}
C(\omega_n) = \left( \frac{B_f}{\gamma_b+B_f} \right) \frac{F_s(0,\omega_n)}{\cosh(k_f d_f)}.
\end{equation}
The Green function at the S side of the SF interface is
\begin{equation}
F_s(0,\omega_n)=\frac{F_s^+(0,\omega_n)+F_s^-(0,\omega_n)}2.
\end{equation}
The symmetric part $F_s^+$ is given by Eq. (\ref{fsplus}). The antisymmetric part is
\begin{equation}
F_s^- = C^-(\omega_n) \cosh\left( k_s [x-d_s] \right),
\end{equation}
with $C^-(\omega_n)$ found from the boundary conditions:
\begin{equation}
C^-(\omega_n) = \left[ \frac{i \gamma\Imag B_f}{A_s |\gamma_b+B_f|^2 +\gamma (\gamma_b+\Real B_f)} \right]
\frac{F_s^+(0,\omega_n)}{\cosh(k_s d_s)}.
\end{equation}
Finally, the order parameter in the F layer is the Fourier transform [see Eq. (\ref{order_parameter})] of
\begin{gather}
F_f(x,\omega_n) = \left[ 1+ \frac{i \gamma\Imag B_f}{A_s |\gamma_b+B_f|^2 +\gamma
(\gamma_b+\Real B_f)} \right] \notag\\
\times \left( \frac{B_f}{\gamma_b+B_f} \right) \frac{\cosh\left( k_f [x+d_f] \right)}{\cosh(k_f d_f)}
\int_0^{d_s} G(0,y;\omega_n) \Delta(y) dy.
\end{gather}


\begin{thebibliography}{99}
\bibitem{T_c_McMillan}
To avoid confusion, we note that Eq. (39) from McMillan's paper\cite{McMillan} determining the critical
temperature of a SN bilayer is incorrect. The correct equation, following from Eqs. (37), (38), (40) and
leading to Eq. (41) of Ref.~\onlinecite{McMillan}, reads $\ln (T_{CS}/T_C) = (\Gamma_S/\Gamma)\, [ \psi (
\frac 12 +\Gamma/2\pi T_C ) -\psi (\frac 12 )]$.
\end{thebibliography}
\end{document}